# Life-Cycles and Mutual Effects of Scientific Communities

Václav Belák, Marcel Karnstedt, and Conor Hayes

Digital Enterprise Research Institute (DERI)

National University of Ireland, Galway, Ireland

first.last@deri.org

**Abstract**

Community effects on the behaviour of individuals, the community itself and other communities can be observed in a wide range of applications. This is true in scientific research, where communities of researchers have increasingly to justify their impact and progress to funding agencies. While previous work has tried to explain and analyse such phenomena, there is still a great potential for increasing the quality and accuracy of this analysis, especially in the context of cross-community effects. In this work, we propose a general framework consisting of several different techniques to analyse and explain such dynamics. The proposed methodology works with arbitrary community algorithms and incorporates meta-data to improve the overall quality and expressiveness of the analysis. We suggest and discuss several approaches to understand, interpret and explain particular phenomena, which themselves are identified in an automated manner. We illustrate the benefits and strengths of our approach by exposing highly interesting in-depth details of cross-community effects between two closely related and well established areas of scientific research. We finally conclude and highlight the important open issues on the way towards understanding, defining and eventually predicting typical life-cycles and classes of communities in the context of cross-community effects.

# 1 Introduction

Claims for scientific progress are often assessed using relatively static citation measures. However, the analysis of the life-cycle of a community provides much greater explanatory power for the progress and potential of a scientific field—for the community itself and external evaluators such as tenure committees, funding agencies, venture capitalists and industry. It can guide funding agencies and tenure committees to make more informed decisions and to identify trends and new funding opportunities. While previous work has examined scientific networks through co-citation and textual analysis, there is relatively little work on analysing the dynamics and behaviours of cross-community behaviours, particularly where closely related communities are competing for scientific, funding and industrial capital.

In previous work [16], we proposed a general road-map for the cross-community analysis of scientific communities. In this work, we elaborate this idea further and present the next results and insights we gained on the way following the proposed road map. Although several recent works deal with the dynamics of communities, we are not aware of any work that discusses general methods for enabling an automated identification and analysis, particularly in the context of cross-community effects. Most of the works are limited to specific algorithms, specific phenomena and specific measurements. Moreover, no works exist that investigate the actual notion of community life-cycles while regarding the aforementioned phenomena as fundamental change points in the life of a community. While some works identify and discuss similar phenomena as we do, they lack in a detailed evaluation of according indicators and explanations.

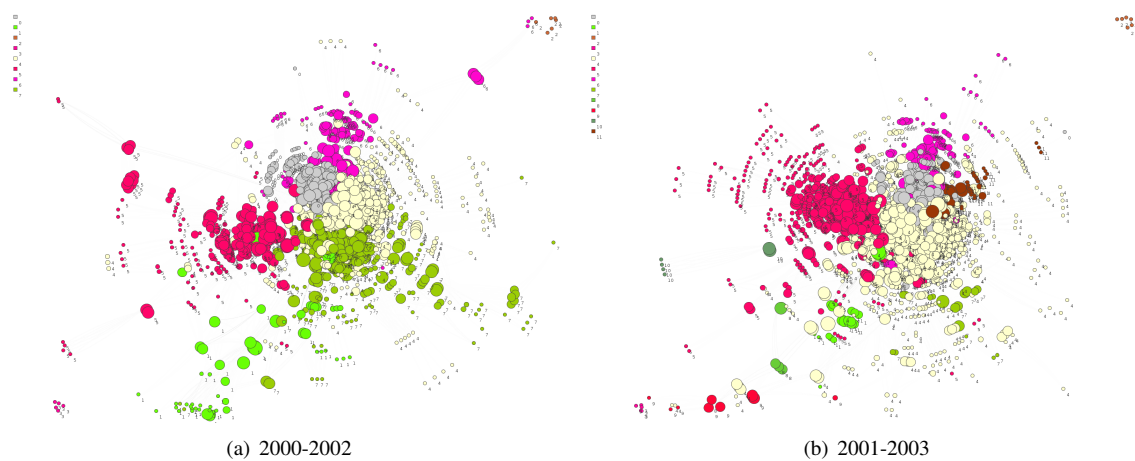

(a) 2000-2002       (b) 2001-2003

Figure 1: The early years of the SW community (red, number 5 on the left of both figures) and its positioning close to the IR field

In this work, we analyse phenomena in the life-cycle and interactions of two closely related areas of scientific research using author-based co-citation network analysis (see, e.g., [11]), supplemented by automatic extraction and investigation of the topics and expertise that form the core of each community. The research fields we chose are Information Retrieval (IR) and Semantic Web (SW). IR research, a well established part of conventional Web search providers, with well-defined methodological and evaluation techniques, is strongly driven by a problem. SW research, on the other hand, focuses on techniques of the next generation Web and is driven by a solution [1]. SW research shows a strong emphasis on breaking the current mould of IR-inspired document-based retrieval in favour of data-focused retrieval. Baeza-Yates et al.'s pithy statement captures the deep cultural divide between the research communities of IR and SW. Both are well-established communities focused on accurate, scalable search methods on the Web. Yet in terms of methodology and research culture, both communities are in many ways orthogonal, with large cores of each discipline indifferent to each other's work. While various sub-communities of these communities are in touch with each other, there are large cores of each discipline that are indifferent to each other, despite their shared goals. As such, the two general research fields promise to reveal a wide range of interesting cross-community effects, influences and interactions. However, we understand this choice as an illustrative example, while our main goal is to develop methods and techniques to analyse cross-community effects between arbitrary research fields.

As a motivating example, Figure 1 illustrates the community structure of the SW and IR fields in the early years of SW (more details of the nature of these figures can be found in Section 4.2). Figure 1(a) shows the newly formed SW community (red, number 5) on the left. By shifting the inspected time interval by just one year (Figure 1(b)), it becomes obvious that the SW community grows and moves closer to IR. But, there are several other interesting insights and phenomena indicated: core IR communities, small evolving and shrinking communities, more structural differentiation inside the IR field than in SW, etc. These first rough insights motivated us to develop sophisticated methods to understand, interpret and at some stage being able to predict all the involved effects and mechanisms.

In this setup, we identify certain phenomena (see Section 3) and trace and explain the evolution of new subcommunities. Interpretation of the results is provided by a specialised visualisation of community dynamics over time, enriched by automatically extracted meta-data and several different measures that promise to be particularly informative. Note that we use the terms communities and clusters interchangeably. In summary, our contributions are:

- We propose techniques to enable scalable analysis of cross-community dynamics. The methodology is not limited to one particular community-detection method and is suited for different relations between individuals.

- We incorporate automatically extracted meta-data, both for enriching the actual analysis as well as enabling new methods of assessing the clustering quality. In other words, we combine topological analysis with topical analysis.

- We discuss and evaluate different methods and measurements to automatically determine community overlap, community relations and specific interesting phenomena.

- We further analyse and discuss *life-cycle measurements* and their suitability for identification and explanation.

We present several outcomes. First, we exemplify how graph analysis and SNA techniques can help to identify and explain very interesting insights into cross-community dynamics, which are not obvious by manual inspection. The combination of visual elements and text enables easier interpretation of the evolution of a scientific research topic, and the motivations of its contributing researchers. Second, we propose new ways to incorporate meta-data analysis. We also demonstrate how meta-data enables a novel way of assessing the quality of detected communities and their dynamics. Third, we build a general, flexible and extendible framework for analysing cross-community dynamics and highlight the suitability of different methods and techniques in this context. However, we also show that this exciting research direction still offers several strands and open issues for future work. The overall goal is to understand, define and eventually predict typical life-cycles and states of communities in the context of cross-community effects.

The remainder of this work is structured as follows. In Section 2, we give a brief summary of the most important related work. In Section 3, we describe the phenomena we expect to identify using the proposed methodology. Based on this, we describe the set of applied techniques and methods in Section 4. Before presenting results gained with these methods in Section 6, we give an overview of the data used for that in Section 5. Afterwards, we discuss the gained results in the context of our overall goal in Section 7, before we finally conclude and indicate future work in Section 8.

## 2 Related Work

Thomas Kuhn introduced the idea of paradigm shift into the lexicon of scientific discourse as a means of explaining how new theories overturn existing theories within a scientific field [17]. Contrary to the conventional view that scientific knowledge is gradually accumulated, Kuhn proposed that science is dominated by periods of stasis ('normal science'), in which empirical evidence is gathered to reaffirm the prevailing theoretical framework. These periods are shattered by periodic abrupt transformations or crises in the theoretical underpinning of a discipline in which a rival paradigm emerges. In Kuhn's view the the reception of a new paradigm necessitates a redefinition of the corresponding science. Opponents of the new paradigm find themselves unable to critique the rival paradigm with respect to the 'normal' theoretical framework and proponents may no longer recognise or value the accepted problems of the old paradigm. Thus, the paradigm shift is typified by a breakdown in scientific communication between proponents and opponents. Communication is only restored by opponents (gradually) accepting the terms of the new paradigm or dying out. Kuhn's analysis tends to focus on well-known dramatic shifts in science such as the juncture between Newtonian and relativistic physics. In this paper, we explore whether similar behaviours occur in what might be termed 'normal science'. We choose two scientific communities, one of which, Information Retrieval, considers itself to be 'normal science' with rigorously defined standards in methodology and evaluation. The Semantic Web community, on the other hand, presents itself as an 'avant-garde' research initiative, addressing (and extending) many of the same problems as IR through Semantic technologies and standards. We make no assumption about the claims either community makes about itself or the other. We are interested in exploring the communication behaviours that occur between them, and whether they can help us to understand how new knowledge and ideas are created and disseminated.

We picked up this idea before in a position paper [16]. In that work, we gave an introduction of the parallels between Kuhn's observations and the effects and phenomena we expect in the interplay of scientific communities. Further, we proposed a road map to analyse and understand these phenomena. The work in hand presents the insights and experiences we gained from the next steps on this proposed road. We believe that the phenomena analysed in this work, which are based on the phenomena described by Kuhn, can be found in many typical life-cycles of scientific communities. We propose and assess methods that help to understand and eventually predict these life-cycles.

Recently, the idea of analysing the dynamics of communities, in contrast to former static community analysis, gained attention. There are several works dealing with this issue. Most of them follow the idea of using snapshots of the underlying network graph from different points in time. Communities found in these snapshots are compared over time and the development of communities is deduced and investigated, such as in [12]. That work is very close to our idea of understanding the life-cycles of communities, but does not investigate a similarly rich set of indicators and automated methods. In analogy to our work, the authors base their analysis on non-overlapping communities. While [20] also investigates the time dependence of communities and community evolution, it rather focuses on overlapping communities. This is expected to be a crucial requirement for domains where individuals are usually members of a large number of different communities. We follow the ideas of these works and develop them further towards the general understanding of typical community life-cycles, the underlying mechanisms and reasons for phenomena aligned with Kuhn's observations. In contrast to former work, we aim for a general methodology supporting arbitrary community detection algorithms – [8] provides an exhaustive overview of existing approaches. Further, we investigate in more detail potential life-cycle measures regarding their potential to indicate and explain typical life-cycles and phenomena.

[26] also deals with the dynamics of communities, but proposes an interesting alternative approach based on a graph-colouring problem. However, in contrast to our work, they rather focus on the dynamics of single individuals and how they switch between communities over time. Moreover, several of the assumptions (such as separated communities not connected among each other) do not apply in our case. We focus on a large-scale analysis of the communities themselves on macro level. The aim is to understand effects between different communities and general research fields, rather than effects between communities and individuals. However, an extension to the micro level of single community members is one of the upcoming steps.

A crucial problem with the snapshot-based approach is the choice of the underlying time periods, which can have significant influence on the gained insights. Too small periods may provide too small and separated communities to identify cross-community effects. On the other hand, too large periods may obfuscate these effects. Just recently some works discuss this in more detail [24, 6]. A related problem is the question on how to map one community between snapshots, which is required to determine the community's evolution. This becomes even more complicated with the use of static community detection, where communities for each snapshot are determined independent from other snapshots. A promising alternative is to use evolutionary clustering [5, 18]. In this approach, the community detection at a certain point in time is influenced by the community structure in former times, aiming at community structures that are more stable over time. We designed our framework to be adaptive and extensible in both directions. While in this work we focus on a fixed time period and existing static community-detection methods, we plan to investigate evolutionary clustering in future work. Related approaches can also be found in the context of data-stream analyses inspecting evolving clusters and change points, e.g., [25].

Hayes et al. [15] examined user and topic drift in communities of bloggers, introducing the notion of blog author entropy. The focus of this work was on examining the relationship between topic drift and author entropy. Unlike the work presented in this paper, clustering was carried out on blog topics rather than on the blog link network. However, this work inspired the topical parts of our analysis.

## 3 Cross-Community Effects and Phenomena

As mentioned in Section 2, the phenomena described by Kuhn [17] highly motivated the research presented in this work. In [16], we initially discussed his observations in the context of cross-community effects in scientific research and described the phenomena we expect. However, a paradigm shift as Kuhn describes it is something very significant, close to a "revolution in science". We investigate closely related phenomena, but of less dramatic importance and with smaller influence on the research field. Thus, in this work we use the term *community shift*. A particularly dominant and huge community shift is what we eventually understand as a paradigm shift.

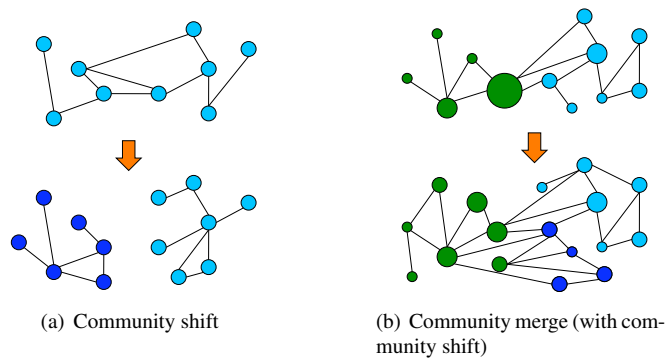

(a) Community shift    (b) Community merge (with community shift)

Figure 2: Community shift and community merge as possible phenomena

Figure 2(a) illustrates what we call a community shift. The upper part shows co-citation relations between different authors at a specific point in time. Over the time, a sub-community detaches from its original community (lower part of the figure). This means, authors from both communities are not cited together any more, with ongoing time the sub-community splits. Figure 2(b) illustrates the opposite of this, which we call a *community merge*. Over the time, the communities approach each other, represented by more and more edges between the members. This can lead to closely related communities or even to a merge into one larger community. For some communities, we even expect a combination of community shift and merge. This means, from one large community only a sub-community approaches the originally separated one—whereby in parallel detaching from its original field of research. We indicate this in the figure by the different colours of the nodes. Note that the effects are illustrated in a rather dramatic manner. We expect these phenomena to usually occur alleviated, which is one of the reasons for using the introduced new terms to describe them.

Another interesting phenomenon described by Kuhn is a paradigm articulation, which refers to the maturing process of a community resulting in different groups specialising on sub-topics. Naturally, such an effect cannot be analysed solely on the basis of the network structure. In this case, the benefits of enriching the topological analysis by topical analysis become obvious. Consequently, we call such a phenomenon *community specialisation*. Similar, a community that stays structurally stable over time might change its topical focus. Again, this can be analysed only if we incorporate topical analysis. Consequently, we call this a *community topic change*.

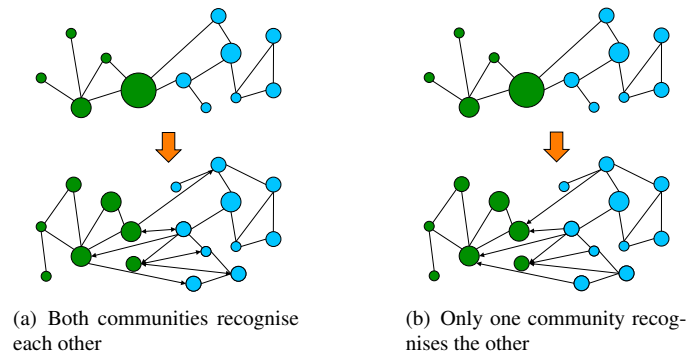

(a) Both communities recognise each other

(b) Only one community recognises the other

Figure 3: Possible "social" and "non-social" behaviour

This is only a selected subset of all phenomena that might occur and that we can imagine. We chose these phenomena due to their close relation to Kuhn's discussion. Moreover, we expect these to occur very frequently, to be particularly obvious and to be clearly indicated by special measures. As we especially regard cross-community effects, it is very important to identify if communities (regularly) exchange members or maybe "feed" other communities. These are the actual aspects that this work focuses on. If it works for the chosen phenomena, it likely forms a suitable basis for other phenomena as well. Thus, later on we are going to extend the set of analysed phenomena. This will involve different phenomena (which by now we might not even think about) as well as different types of networks and relations (directed citations, bibliographic coupling—see [7] for an illustrative comparison to the co-citation measure, co-authorship, ...). As an example, Figure 3 illustrates possible behaviour in the context of how communities recognise each other. One of the communities might show a "non-social" behaviour, simply neglecting the existence and development of new communities. In this case, we expect only one community to cite the other. In contrast, a healthy development would be observed if both communities increasingly cite each other over time. Figure 3 illustrates that difference by using directed edges that indicate the direction of citations. This leads to other important questions, such as what actually defines social behaviour and how to detect it. The framework proposed in this work provides an extensible basis for enabling such further analyses.

Other interesting effects include shifts of visibility between the actors of evolving communities, growth, death and health of communities. Especially for new and rapidly evolving communities like the field of SW, the only work that is "visible" to other communities might be a very fundamental and ground-breaking contribution by one of its founders. For the SW, one can think of Tim Berners-Lee famous work [2], which is seen as the initial work founding that community. We indicate such a visibility by using differently sized nodes in the figure. Over time, more works "appear on the horizon", the coastline of the community becomes visible and more islands are cited.

This results in a shift of visibility between the actors of the evolving community. In contrast, certain fields might tend to cite only the "tall" figures visible, even if the community that these figures belong to matures. Analysing such effects requires a look at the micro level of individuals and again the use of different types of relations, such as directed citations.

We understand the work presented here as a fundamental step towards analysing and understanding the full spectrum of possible phenomena and cross-community effects. We show that the proposed methods are suited for this goal and further highlight techniques for an appropriate indication and explanation of the selected phenomena.

# 4 Methodology

Automated identification of the phenomena introduced in Section 3 is a challenging task and a suitable set of methods, measures, and analytics is required. In this section, we motivate and explain the according techniques used in our work. This methodology was developed with a set of certain requirements arising from the nature of the problem:

1. We expect the dynamics of the data set represented by snapshots of several consecutive time-steps.

2. Communities have to be identified in the network in each time-step.

3. Authors (nodes in general) have to be uniquely identified *among all* time-steps.

4. For topical analysis, meta-data (i.e., topics) has been assigned to nodes in the network.

## 4.1 Community Detection and Tracking

We identified communities in each period using three popular community-detection algorithms, which we denote as:

- Infomap [23]

- Louvain [3]

- WT [27]

Whereas WT and Louvain are both based on modularity [19] optimisation, i.e., the topological feature of clustering, the Infomap reveals the community structure according to the information flow in the network modelled as a random walk. Both approaches make sense in our setup. The topological approach inspects the co-citation structure from a rather static point of view, whereas the approach using information flow inspects the dynamic flow of topics between the authors. The underlying co-citation network can be interpreted in both ways.

We chose particularly these algorithms, because all of them are able to operate over weighted networks, they scale up to the size of the analysed network and for each an implementation is publicly available. Moreover, they produce non-overlapping communities. Therefore it is possible to easily visualise them. However, the requirements listed before can be fulfilled by a wide range of community-detection algorithms. Supporting overlapping communities is possible as well, but would require some modifications of the measures presented in the following.

Communities are identified independently for each time step and thus it is necessary to find the counterpart of each community in a subsequent time-step. We track the community throughout the time by means of the highest overlap measured by the Jaccard coefficient [21, 14]. The $i$-th community mined in time $t$, i.e., $c_i^t$, is matched according to the highest Jaccard coefficient value among all communities $C^{t+1}$ mined in time $t+1$:

$$match(c_i^t) = \underset{c_j^{t+1} \in C^{t+1}}{\arg\max} \frac{|c_i^t \cap c_j^{t+1}|}{|c_i^t \cup c_j^{t+1}|} \qquad (1)$$

In case two communities $c_i^t$ and $c_k^t$ both have the maximal overlap with the same subsequent community $c_j^{t+1}$, the matching is again determined by the Jaccard coefficient value. The community that has the higher overlap with $c_j^{t+1}$ is matched, the other community is then matched to the subsequent community with the second-highest overlap. Note that this method of matching would need a modification in case of overlapping communities. [21] proposes a solution for this.

Besides matching of communities between subsequent time-steps, other types of relations among communities like important ancestors or descendant are interesting as well. These relations can be defined as a modification of the Jaccard coefficient, where the overlap is relative to either the latter or the former community:

$$ancestor(c_i^t, c_j^{t+1}) = \frac{|c_i^t \cap c_j^{t+1}|}{|c_j^{t+1}|} \qquad (2)$$

$$descendant(c_i^t, c_j^{t+1}) = \frac{|c_i^t \cap c_j^{t+1}|}{|c_i^t|} \qquad (3)$$

This allows us to figure out important ancestors of community $c_j^{t+1}$ (Eq. 2) or important descendants of community $c_i^t$ (Eq. 3).

## 4.2 Visualisation

To easily compare and inspect the state of the network in different time periods, we have implemented a visualisation tool in Java using JUNG.[1] We had two reasons for implementing an own visualisation tool: firstly to preserve positions of nodes that have already appeared in previous time periods, and secondly to preserve colours of nodes, which denote the affiliation of the node to its cluster. We have not found any existing tool that implements these features. As a layout, we chose the Fruchterman-Reingold force-driven algorithm [10] as implemented by JUNG, which naturally clusters the nodes. This tool is independent of the clustering method—as far as it produces non-overlapping communities—and it provides a quick insight into the life of communities. Figure 1 consists of two snapshots of the network obtained by our tool. The size of a node represents the logarithmically scaled betweenness of that author. A comparison of two subsequent snapshots like Figure 1(a) and Figure 1(b) enables an observer to state a hypothesis about the cross-community dynamics, which can then be supported or rejected by overlap and other measures. Note that we omit the edges from the graphs for illustrative reasons.

---
[1] Java Universal Network/Graph framework, see http://jung.sourceforge.net.

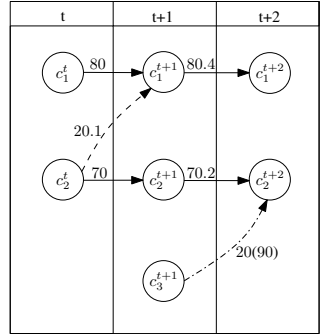

Figure 4: Example of a community evolution diagram. Solid lines denote *match* and dashed lines *ancestor* relationships, while dashed-dotted lines denote a community merge. Annotations of solid and dashed lines express values of *ancestor* relations, e.g., $c_1^{t+1}$ is formed by $20.1\%$ members coming from $c_2^t$ and by $80\%$ members from $c_1^t$. Annotations of dashed-dotted lines express *ancestor(descendant)* values, i.e., $90\%$ of members of community $c_3^{t+1}$ moved to $c_2^{t+2}$, thus forming $20\%$ of all of its members. Note that all values are presented as percentages rounded to the nearest tenth and that the sum of all of ancestors need not to be $100\%$, since some of the members may be completely new.

Particularly, we make use of these measures to generate diagrams of descendants or ancestors of any community of interest using the Graphviz package.[2] This is an automated process. However, to identify and isolate interesting relations, some parameters have to be set by hand (minimal weight of displayed edges, included communities, ...), which involves some experimentation. Otherwise, long-living communities will result in very dense networks of their relatives, making it hard to identify anything. Note that for presentation purposes we redraw these diagrams manually. An example is depicted in Figure 4.

## 4.3 Topic detection

Using solely a community structure obtained by a community-detection algorithm, we cannot explain all the mutual inter-reactions of communities. A more detailed and expressive picture of a community life-cycle can be obtained by assigning meta-data to communities. For example, these meta-data can be summarised to "name" communities. In order to identify the community topics, we mined keywords using NLP techniques [4] from the abstracts or full-texts for almost $70\%$ of the underlying articles. In addition, author-provided keywords for $10\%$ of the articles were extracted. See Section 5 for more detailed numbers. All the keywords were tokenised and stemmed [22]. Every author was then assigned with a set of keywords associated with his articles. Keywords were ranked by TF-IAF measure, i.e., analogous to TF-IDF, but with keywords assigned to authors instead of documents. To determine the keywords for a given time period, we selected documents written by an author in that time and then assigned the keywords from only those documents to that author. In case a document had more than just one author, the keywords were assigned to all of them. As a result, each author $a$ was described by a bag-of-words vector $k_a^t$ for each time period $t$.

---

[2]See http://www.graphviz.org/.

However, with keywords assigned to authors one still cannot figure out the overall topic of a cluster. In content analysis, the overall cluster topic is usually characterised by its centroid. In contrast with usual content clustering methods, where these centroids are a result of the clustering process (e.g., k-means) itself, in our case the clusters are mined independently. The topical centroids are thus derived from the keywords of all cluster members, according to the standard formula for centroid computation, which reads:

$$centroid(c) = \frac{\sum\limits_{a \in c} k_a}{|c|} \tag{4}$$

Centroids were used for both computation and interpretation purposes. Namely, the measures discussed in the next section are based on (dis)similarity obtained as a standard cosine distance between two centroids. The total number of keywords, i.e., the dimension of $k_a^t$ vectors, usually exceeded several thousands. Therefore, for interpretation purposes it was necessary to consider only the highest-ranked ones. Hence, the interpretation of a cluster topic was derived from its 20 highest-ranked keywords. However, this led sometimes to very rare keywords to be ranked highest, while the general yet informative keywords were discarded by the TF-IAF ranking. This is because TF-IAF ranks keywords according to their uniqueness in the corpus (IAF) as well as their frequency (TF). Hence, in addition to the 20 highest-ranked keywords, we also considered the 20 most frequent keywords (TF) of the cluster. General but still informative keywords like "web", "information", or "retrieval" had usually high TF but small IAF values. We will refer to the union of these two sets of keywords as *characterising keywords*.

## 4.4 Measures

Applying the described overlap measure results in a high number of overlaps between many different communities. Thus, the amount of different mutual interactions is huge and unmanageable, even with only a couple of communities during few years. The natural step to uncover potentially interesting events or even to predict them is to apply more specific measures or to use the simple ones in combination. For this purpose, we developed two categories of additional measures:

1. community life-cycle measures

2. community topic evolution measures

The purpose of the first category is to measure and explain the state and the evolution of the community. The second category of measures is focused on revealing interesting topical changes of a community discussed in Section 3, e.g., emergence of a new community topically distinct to its ancestor (community shift). In general, we combined both topological and topical (i.e., content) measures to obtain a deeper and more informative picture of a community life-cycle.

### 4.4.1 Community Life-Cycle

From the structural perspective, clusters were described by *size* $\mathcal{S}$, average *vertex betweenness* $\mathcal{B}$, *author entropy* $\mathcal{A}$ and *relative density* $\rho$. We chose this set of measures because they had been successfully used before in the literature and promised to be particularly informative. Author entropy has been defined and explained by Hayes et al. [15]:

$$\mathcal{A}(c^{t+1}) = -\frac{1}{\log|C_o^{t+1}|} \sum_{c_o \in C_o^{t+1}} \frac{|c^t \cap c_o|}{|c^t \cap A^{t+1}|} \log \frac{|c^t \cap c_o|}{|c^t \cap A^{t+1}|}, \quad (5)$$

where $A^{t+1}$ is the set of all authors in time $t+1$ and $C_o^{t+1}$ is the set of clusters in time $t+1$ containing authors of $c^t$. $\mathcal{A}$ measures how much the authors of $c^t$ are dispersed among other clusters in a subsequent time-step. If the authors are equally dispersed among subsequent communities, $\mathcal{A}$ will approach 1, whereas if all the authors remain in the same community, $\mathcal{A}$ will approach 0.

Relative density $\rho$ is defined as the ratio between intra-cluster degree and its total degree:

$$\rho(c) = \frac{\sum_{e \in E_{c_i}} w(e)}{\sum_{e \in E_{c_a}} w(e)}, \quad (6)$$

where $E_{c_i}$ is the set of all internal edges of cluster $c$, $E_{c_a}$ is the set of all edges incident to cluster $c$ and $w$ is a function assigning a weight to each edge. $\rho$ is a *local* measure of cluster quality. As a community is usually understood as a subgraph with more intra- then inter-cluster edges, we chose this measure to investigate the level to which the community is structurally shaped. In case of self-referential communities, i.e., those ones without any edge to any other community, $\rho$ will be 1, whereas a very open and ill-shaped community will have values near to 0.

From the topical point of view, the *topic drift* $\mathcal{T}$ and *cluster content ratio* $\mathcal{H}$ as they are discussed in [15] were used.[3] $\mathcal{T}$ is the cosine distance between centroids of the same cluster in two subsequent time periods. $\mathcal{H}$ is the ratio of intra- to inter-cluster similarity:

$$\mathcal{H}(c) = \frac{\mathcal{I}}{\mathcal{E}} = \frac{\frac{1}{|c|}\sum_{a \in c} cos(a, centroid(c))}{cos(centroid(c), centroid(A))}, \quad (7)$$

where $\mathcal{I}$ is the average similarity between the cluster's authors and its centroid and $\mathcal{E}$ is the similarity between the cluster's centroid and the centroid of the entire network ($A$ is the set of all authors).

### 4.4.2 Community Topic Evolution

The interesting events in a community life-cycle are characterised by a change in the structure of a community and an accompanying change of its topic. However, communities naturally evolve, so it is necessary to discriminate interesting events from uninteresting ones. As the total amount of potentially interesting events is usually large, it is necessary to employ some automated technique to mine only the events of potentially high interest. Any measure supposed to do that has to consider both topical and structural changes. Three measures expressing those changes have already been defined: *ancestor* (Eq. 2), *descendant* (Eq. 3) and author entropy $\mathcal{A}$ (Eq. 5). The change of topic can be expressed as a dissimilarity *dissim* defined as a complement to cosine distance. By combining these two kinds of measures, it is possible to construct various measures that reveal the phenomena we are interested in. We simply combine them by multiplication, because then the product remains within $[0, 1]$. This enabled us to specify a simple threshold pruning unwanted selections.

---
[3] Note that for the sake of clarity we changed the name of $\mathcal{H}$ from *cluster quality* as originally termed to *cluster content ratio*.

In case of a *community shift*, we are interested in a newly emerged community significantly different from its ancestor. On the contrary, a community shift combined with merge, *community shift/merge*, can be detected as an event when one community merges with a topically different community. The merge can be expressed as a *descendant* relationship, especially if $descendant(c_i^t, c_j^{t+1}) \to 1$. Thus, the used formulae for community shift $\mathcal{P}_\mathcal{S}$ and community shift/merge $\mathcal{P}_{\mathcal{S}/\mathcal{M}}$ for $i \neq j$ read:

$$\mathcal{P}_\mathcal{S}(c_i^t, c_j^{t+1}) = dissim(c_i^t, c_j^{t+1}) \times ancestor(c_i^t, c_j^{t+1}) \tag{8}$$

$$\mathcal{P}_{\mathcal{S}/\mathcal{M}}(c_i^t, c_j^{t+1}) = dissim(c_i^t, c_j^{t+1}) \times descendant(c_i^t, c_j^{t+1}) \tag{9}$$

$\mathcal{P}_\mathcal{S}$ and $\mathcal{P}_{\mathcal{S}/\mathcal{M}}$ express the relatedness of one community to another in terms of their structure and difference in terms of the topic. The more one community is formed by members of another community and the more these two communities differ in topic, the higher these values will be. Naturally, we expect a significant difference between the sizes of communities $c_i^t$ and $c_i^{t+1}$ in both phenomena, i.e., the shifting community is usually smaller then its ancestor, and the merging community is usually smaller then its descendant. Thus, we used a threshold of 0.5 with these measures. With this choice, for the maximum value of $dissim(c_i^t, c_j^{t+1}) = 1$ only shifts or merges with at least 50% membership overlap are selected.

Another interesting change in the life of a scientific community is when it changes its topic while preserving its structure. To detect those cases, we defined the *community topic change* $\mathcal{P}_\mathcal{C}$ measure as:

$$\mathcal{P}_\mathcal{C}(c_i^t) = dissim(c_i^t, c_i^{t+1}) \times (1 - \mathcal{A}(c_i^{t+1})), \tag{10}$$

where $\mathcal{A}$ is the author entropy as defined in Eq. 5. $\mathcal{P}_\mathcal{C}$ measures the change of the topic of the cluster $c_i$ between subsequent time-steps. This discriminates cases where the cluster has very weak structure, since then the entropy will be high. As we usually observed an entropy greater than zero, we chose a threshold of 0.3.[4]

Even with the use of thresholds, we detected events associated with very small communities, i.e., with size $\mathcal{S} < 5$. Hence, we decided to use a minimal value of overlap in addition. Only changes associated with a minimal

---
[4]The average entropy and its variance for Louvain clusterings were $\langle\langle \mathcal{A}_L \rangle\rangle \doteq 0.4$, $\sigma^2_{\mathcal{A}_L} \doteq 0.1$. For Infomap clusterings these values were $\langle\langle \mathcal{A}_I \rangle\rangle \doteq 0.44$, $\sigma^2_{\mathcal{A}_I} \doteq 0.14$.

overlap of 5 authors were considered for deeper analysis.

### 4.4.3 Inter-Camp Dynamics

All the measures discussed so far are general and are independent of the actual research field a community belongs to. However, we are particularly interested in cross-community dynamics between IR and SW camps. Thus, it was necessary to determine for each analysed community to which of the fields it belongs. Hence, we identified an SW-related community as one having at least one of the (stemmed) keywords "semant", "ontolog" or "rdf" among its characterising keywords. Similarly, an IR-related community was identified as a community having at least one of the "ir" or "retriev" among its characterising keywords. These keywords were chosen according to the most frequent patterns mined from publications in both research field. An event detected by any of the community topic evolution measures was then considered as *inter-camp dynamics* if both SW- and IR-related communities were involved in it. Presupposing both IR and SW camps are rather separate (see Section 1), communities with both IR- and SW-related themes should be distinguishable by high average betweenness. Therefore, the communities featuring inter-camp dynamics and high betweenness values were of our particular interest, as this promised to identify communities connecting these two camps.

## 4.5 Clustering Assessment

| algorithm | $\langle\langle\mathcal{Q}_o\rangle\rangle$ | $\sigma^2_{\mathcal{Q}_o}$ | $\langle\langle\mathcal{H}_o\rangle\rangle$ | $\sigma^2_{\mathcal{H}_o}$ |
|---|---|---|---|---|
| Louvain | 0.56085 | 0.01489 | 2.01912 | 4.33 |
| Infomap | 0.74136 | 0.00397 | 1.74818 | 1.86672 |
| WT | 0.62246 | 0.01321 | 1.59136 | 4.45796 |

Table 1: Overall average and variance values of cluster content ratio $\langle\langle\mathcal{H}_o\rangle\rangle$ and modularity $\langle\langle\mathcal{Q}_o\rangle\rangle$ for clusterings obtained by used algorithms

Interestingly, the cluster content ratio $\mathcal{H}$ can be used to assess the quality of the clustering. $\mathcal{H}$ expresses how well the cluster is shaped in terms of its topic. The average value of $\mathcal{H}$ for each clustering can therefore be used to measure the ability of a community-detection algorithm to produce topically coherent clusters, which are mutually topically distinct.

From the topological point of view, modularity $\mathcal{Q}$ is probably the most popular quality function today [8]. It is based on the idea that the edge density inside a community should be higher than the density of any subgraph of a random graph (i.e., without any community structure). Therefore, it compares the clustering with a *null model* of the network without any community structure. Usually the null model is obtained by randomly rewiring the edges of the graph while persisting the degrees of its vertices. In contrast to the relative density $\rho$, modularity is a *global* quality function of clustering. Therefore it can be conveniently used for assessment of clustering obtained from different algorithms.

Figure 5 illustrates the modularity $\mathcal{Q}$ of the clusterings obtained with each investigated algorithm and the average cluster content ratio $\mathcal{H}$ of the corresponding clusters per year. Overall average values over all time periods are shown in Table 1. Surprisingly, the two algorithms based on modularity maximisation produced clusterings of lower modularity than Infomap, which performed the best in that regard. On the other hand, Louvain had the highest average cluster content ratio. However, the variance indicates the higher volatility of this feature of Louvain clusters. Both Louvain and Infomap clusters generally had a high cluster content ratio with $\langle\langle\mathcal{H}\rangle\rangle > 1$ for every year, which means these clusters were topically well-shaped. WT performed worse in that regard between 2000 and 2004. The high overall average cluster content ratio of Louvain clusterings $\langle\langle\mathcal{H}_o\rangle\rangle \doteq 2$ shows that the centroids of these clusters were twice as similar to the contained authors as they were to the centroid of the entire network. In other words, the members of these clusters were twice as similar to each other as to the rest of the network. Comparing the overall average modularities and cluster content ratios, it is apparent that WT performed worse than Louvain wrt. $\mathcal{H}$ and worse than Infomap wrt. $\mathcal{Q}$. Since we found much less interesting inter-camp dynamics between clusters obtained by WT and because its performance was dominated by other algorithms, we do not discuss communities detected by this algorithm in the results section.

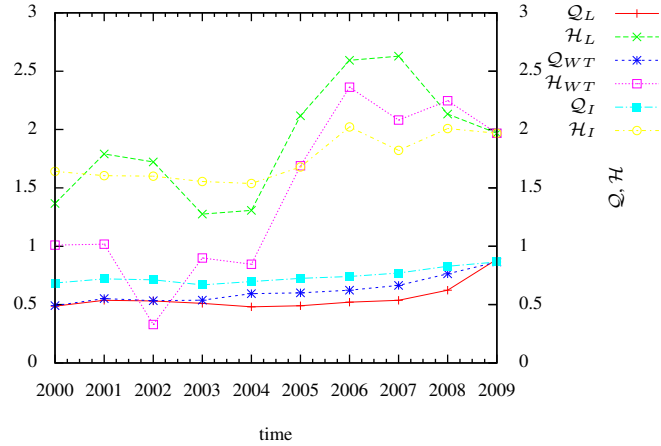

Figure 5: Modularity and average cluster content ratio per year. Subscripts $_L, _{WT}$, and $_I$ denote Louvain, WT, and Infomap algorithms, respectively.

## 5 Data Set

In order to build networks of actors from the SW and IR fields, we first picked a set of major conferences from each field. For the time being, we chose only conferences, not journals. The reason for this is that in computer science conferences play a much more important role than in other disciplines. Particularly, if we want to identify young and evolving communities, conferences are better suited than journals. For the SW field, we chose the International, Extended (former "European") and Asian Semantic Web Conferences (ISWC, ESWC, ASWC) and co-located workshops. In addition, we considered all work from Tim Berners-Lee, as the generally understood founder of SW. The IR field is represented by the Special Interest Group on Information Retrieval conference (SIGIR), European Conference on Information Retrieval (ECIR), the International Conference on Information and Knowledge Management (CIKM), Text Retrieval Conference (TREC) and all co-located workshops.

We selected all publications available for these venues from DBLP[5] as seeds. DBLP provides meta-data (authors, years, ...) for a large set of publications, cleaned in a time-consuming pre-processing phase. We selected all according publications starting from the year 2000, which we see as the "official birth" of the SW field (at least, with respect to the existence of according conferences and workshops and the availability of DBLP data). Unfortunately, DBLP does not provide citation information, crucial for our approach. Thus, we used crawlers to fetch this information from appropriate Web sources. As this inevitably results in problems about author and title disambiguation and "dirty" data, we had to spend a large amount of time for developing data-cleaning methods. To avoid the inclusion of too much dirty data, we mapped citing articles back to DBLP and extracted the required meta-data only from there. Consequently, we could not add all citations that the Web resources offered. On the other hand, we filtered out all references to documents that are not published on main scientific conferences (reports, theses, ...).

| time period | authors | edges |
|---|---|---|
| 2000–2002 | 1459 | 66039 |
| 2001–2003 | 1906 | 87520 |
| 2002–2004 | 2211 | 107499 |
| 2003–2005 | 2468 | 120471 |
| 2004–2006 | 2776 | 141093 |
| 2005–2007 | 3062 | 134132 |
| 2006–2008 | 3002 | 102928 |
| 2007–2009 | 2190 | 44461 |
| 2008–2009 | 1113 | 13340 |
| 2009 | 83 | 159 |
| average | 2027 | 81764.2 |

Table 2: Statistics of the used data set

With this method, we gathered a set of 5772 authors over all years. Using the crawled citation information, we built co-citation networks of these authors. This resulted in 817642 edges in total. Table 2 shows the exact values for the different time periods. We tested several different time periods and found that the best suited networks are based on a three-year period, with an overlap of two years (2000–2002, 2001–2003, ...). As we gathered the data in 2009, the networks for the last years (2009 and 2008) are rather small and sparse, due to the "latency" with which citation links become available. Consequently, we focused on the earlier years in our analysis.

The total number of included articles is 39314. For extracting meta-data to enable the topical analysis, we were able to scrape 22975 abstracts and 3740 full texts. Thus, the total coverage by content was nearly 70%. Further, we scraped 18313 author-provided keywords for 4102 distinct articles—i.e., the coverage by these high-quality keywords was nearly 10%. From the scraped abstracts and texts, we mined 263742 additional keywords.

Intuitively, to analyse dynamics of the same individuals over time, we chose an author-based approach rather than a document-based approach. We constructed co-citation networks where the weight of an edge refers to the number of occurred co-citations in each analysed time period. A co-citation between an author $A_1$ of document $D_1$ and another author $A_2$ of document $D_2$ occurs if we find a third document $D_3$ citing $D_1$ and $D_2$, where $D_1$ and $D_2$ are published in the inspected time period. We chose to restrict the appearance of $D_1$ and $D_2$ to the given time period, rather than $D_3$, for several (in parts practical) reasons. The discussion of this non-trivial issue is out of the scope of this paper. We used the networks weighted in that way as input for the Louvain algorithm, as the available implementation did not allow to use normalised values. However, as a normalised weighting scheme is agreed in the literature to produce better results, we used CoCit scores [11] as input for Infomap. In contrast to Louvain, the available implementation of Infomap supports the usage of floating-point values.

---
[5] http://www.informatik.uni-trier.de/ ley/db/index.html

# 6 Results

The community topic evolution and life-cycle measures were used for analysis of the aforementioned data set. Our aim was to assess the suitability of the methodology to detect and explain the cross-community phenomena like community shift or topic change. We expected insights into the ability of both types of measures to reveal, characterise and explain these phenomena. This section presents a selection of the most interesting cases identified by community topic evolution measures and uses the life-cycle measures to explain them. These cases are structured according to the measure that was mainly used for their identification—even though in some, especially in more complex, cases more than just one measure was helpful. During our experiments and analysis, we made use of all the introduced techniques. However, we present them only where they provide interesting insights and in an appropriate way. Thus, we present visualisation, tag clouds, etc. only in some cases. If we present a tag cloud, for the sake of brevity, we provide only a summary of the key topics of the community. Note that life-cycle measures $\mathcal{A}$ and $\mathcal{T}$ are always computed wrt. to previous time step, e.g., $\mathcal{A}$ in time $t$ measures the level of dispersion of users forming the community in time $t-1$. The analysed data sets are overlapping by two years, which is depicted in figures as an interval value, e.g., 2000–2002. But, for the sake of clarity, in the text we refer to the time-slots only by the beginning of the interval, e.g., 2000.

## 6.1 Community Shift

The emergence of Louvain community 26 in 2006 has been identified as an inter-camp community shift with $\mathcal{P}_\mathcal{S} \doteq 0.62$. As Figure 7 illustrates, it was formed by 80% members of community 6 "web information retrieval" and by 20% of community 5 "semantic web". Right after its emergence, the characterising keywords suggest the focus on interdisciplinary topics like "navigation", "personalization" and "semantic web". Under a massive influence of community 15 "semantic web and IR" in 2007, community 26 changed its topic towards "semantic web and business processes". The focus on mainly SW-related topics remained until 2008, where IR-related keywords appeared again among the characterising keywords. This influence is noticeable in Figure 6(b), where the community is positioned nearer to community 15, in contrast to its original position in the bottom-right corner of Figure 6(a). The higher average betweenness $\mathcal{B} \doteq 2261$ in 2007 in contrast with $\mathcal{B} \doteq 1606$ in 2006 suggested

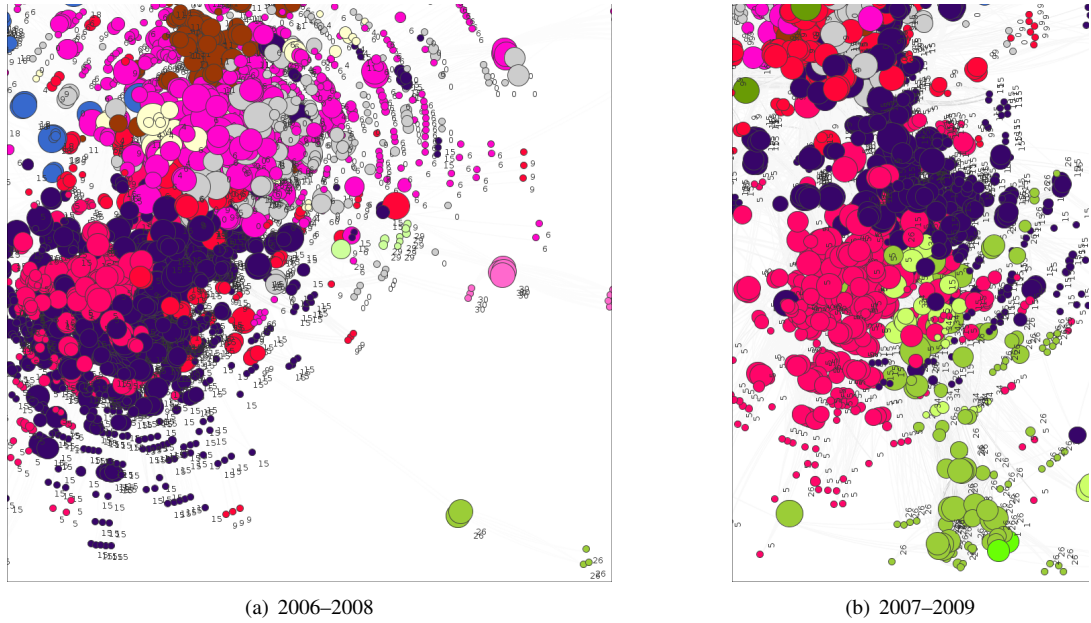

(a) 2006–2008  (b) 2007–2009

Figure 6: Communities 6 "information retrieval" (pink), 5 "semantic web" (red), 15 "semantic web and information retrieval" (violet) and their descendant community 26 (green)

it might have been an intermediary community between SW and IR communities. This is observable in Figure 6, where the sizes of nodes of community 26 are generally bigger in 2006 than in 2007.[6] However, the analysis of characterising keywords did not support this hypothesis, because IR-related keywords appeared among them as late as in 2008. On the other hand, the data set for this time-step is smaller than for the previous one, which might have biased the value of average betweenness. The change of topic towards more SW-related themes in 2007 is expressed by low topic drift $\mathcal{T} \doteq 0.29$ in 2007, while we investigated a rise to $\mathcal{T} \doteq 0.65$ in 2008. This shows the topical stabilisation.

Surprisingly, this pattern repeated in other shifts as well. First, a newly emerged community had a low topic drift, which means it significantly changed with respect to the previous time-step. This improved in further time-steps and as the community was growing, its topic stabilised as well. Community size and topic drift were thus useful for identifying shifts. What seemed in the beginning to be a topically weak community, then grew and grasped its own topical identity.

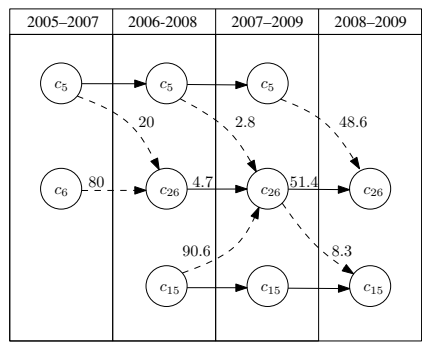

Figure 7: Emergence and the main ancestors of community 26

Shifts may also be interpreted as a *community specialisation* (see Section 3), when new communities with more specialised topics emerge, while the original community becomes eventually smaller. This is the case of Infomap community 9, which started with core concepts around SW and as several specialised communities split off in subsequent time steps, it concretised its topic towards "semantic web services". While the topic drift $\mathcal{T}$ had been very high since the beginning (see Figure 8(a)), the size $\mathcal{S}$ of the community plummeted since 2003, while the cluster content ratio $\mathcal{H}$ started to rise at the same time. Between 2001 and 2004, we identified two community shifts towards more specific SW-related topics from this community. $\mathcal{S}$, $\mathcal{T}$ and $\mathcal{H}$ provided valuable insights into these shifts, as they supported the hypothesis of specialisation: the topically stable community (high $\mathcal{T}$) started to contract (diminishing $\mathcal{S}$), while several distinct communities shifted, which was accompanied by rising content cluster ratio $\mathcal{H}$. Other measures like relative density $\rho$, betweenness $\mathcal{B}$ or author entropy $\mathcal{A}$ did not seem to provide any further insights in this case.

---
[6]We recall that node sizes denote logarithmically scaled node betweenness.

One of the communities shifted from community 9 was community 99 "semantic desktop and personalization", which emerged ($\mathcal{P}_\mathcal{S} \doteq 0.53$) in 2003. The low topic drift $\mathcal{T}$ and relative density $\rho$ and high author entropy $\mathcal{A}$ at the beginning suggest that the cluster was not very well defined during the first two time steps (see Figure 8(b)). But, in 2005 this changed dramatically. Since this time step, community 99 showed high topic drift and relative density. We assume that this is not a coincidence, because in 2006 the main EU project on social-semantic desktop NEPOMUK started.[7] This is a similar pattern to the one observed in Louvain community 26 discussed above.

Community shift measure $\mathcal{P}_\mathcal{S}$ proved to be a useful measure to reveal corresponding phenomena like shifts and specialisation. For their explanation and interpretation, especially topic drift $\mathcal{T}$, cluster content ration $\mathcal{H}$, size $\mathcal{S}$ and relative density $\rho$ were helpful, while the average betweenness $\mathcal{B}$ and author entropy $\mathcal{A}$ did not seem to be very valuable for the analysis of these cases.

## 6.2 Community Shift/Merge

This type of inter-community dynamics seems to be very rare as we have identified only one shift/merge with absolute overlap of 11 authors and $\mathcal{P}_{\mathcal{S}/\mathcal{M}} \doteq 0.91$ between Infomap communities 86 and 0. Table 3 shows both communities were concerned with IR-related topics in general, while each had its specific theme: 86 being focused on "development", "engine" and "system", whereas 0 being focused on "question answering". The interaction of these two and other related communities is illustrated in Figure 9. The community 86 emerged as a descendant of communities 49 "data mining", 4 "cross-language IR" and 43 "web information retrieval", but this is not the phenomena we were investigating in this case. The merge of this community in the following time-step was of our particular interest. It merged with community 0 by 90.9% authors moving to 0 in 2003. Relative density $\rho \doteq 0.47$ and cluster quality $\mathcal{H} \doteq 1.91$ suggest that community 86 was topically coherent, but structurally rather weak. In spite of the strong topic, the community 86 thus dissolved to its related community 0.

As we have identified only one significant shift/merge, it is not possible to generalise the suitability of any life-cycle measure. The community shift/merge measure requires the whole topically distinct community to merge with another one. This is may be a too strong requirement, as it is possible that more often just smaller groups of bigger communities split off and then merge with topically distinct communities. This further suggests to investigate a special *community split* measure.

---

[7]See http://nepomuk.semanticdesktop.org.

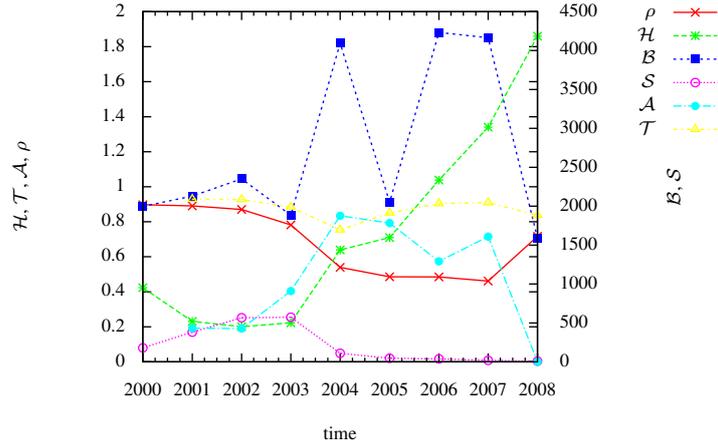

(a) Life-cycle measures of community 9

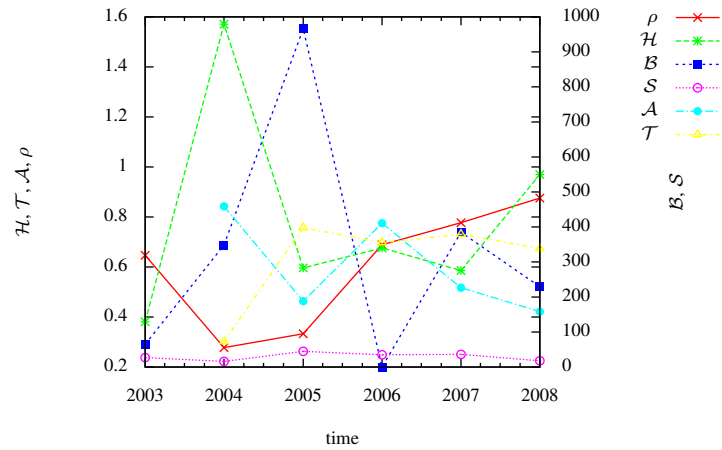

(b) Life-cycle measures of community 99

Figure 8: Life-cycle measures of Infomap communities 9 and 99

## 6.3 Community Topic Change

Community topic change is a significant change of theme of a structurally stable community. One of these changes we identified ($\mathcal{P_C} \doteq 0.58$) was an inter-camp change of the topic of Infomap community 54 between 2005 and 2006. Tag clouds presented in Table 4 show clearly that the focus of the community moved from knowledge management and information extraction ("ie") towards knowledge querying and the semantic web. A zero author entropy $\mathcal{A}$ in 2006 (see Table 5) suggests that this change might have been caused by completely new members joining the community. The main ancestors (see Figure 10) in 2006 were communities 29 "ontologies and SW", 70 "ontologies and folksonomies" and "semantic web services". However, 34.5% of all the authors of community

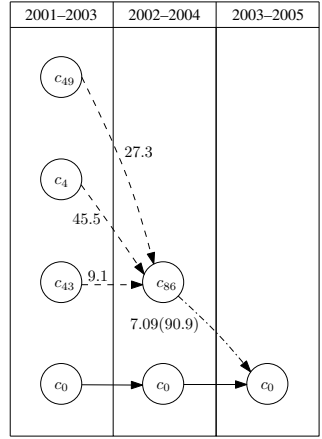

Figure 9: Evolution and the main ancestors of community 86

| community | characterising keywords |
|---|---|
| $c_{86}^{2002}$ | intuitive, development, ir, retrieval, control, implemented, describing, high-dimensional, reducing, engine, execution, advanced, information, system, multi-dimensional, image, usin, accurate, time, precise, features, queries, service, dataset, document, analysis, large, structure, cluster, and, web, processing |
| $c_{0}^{2003}$ | resolution, evaluation, passages, architecture, question, qa, patterns, definitions, development, trec, mit, candidates, linguistic, retrieval, answering, system, analysis, javelin, modules, advanced, methods, science, information, approaches, processing, using, computer, language, techniques |

Table 3: Characterising keywords of Infomap communities 0 and 86. The keywords are listed in their original, i.e., non-stemmed, form.

54 in 2006 were new, i.e., they did not come from any previous community.

Finally, the last case described is Louvain community 15, which was detected by both community shift $\mathcal{P}_\mathcal{S}$ and community topic change $\mathcal{P}_\mathcal{C}$ measures. The emergence and its evolution are depicted in Figure 12. This community first emerged as a descendant of community 4 "information retrieval" with a specific topic "cross-language IR", which has been detected as a shift ($\mathcal{P}_\mathcal{S} \doteq 0.55$). In 2003, this community was under a massive influence of community 5 "semantic web" with 53.1% of its members in 2003 coming from that community. In the same year, the change of community topic ($\mathcal{P}_\mathcal{C} \doteq 0.31$) occurred and the community began to be concerned mainly with SW until 2005, when it was characterised by IR-related keywords as well. As Figure 11(a) depicts, it consisted of two parts in the beginning, which then merged and the whole community moved right between

| community | characterising keywords |
|---|---|
| $c_{54}^{2005}$ | organizational, kms, sw, capturing, environment, working, ie, acquisition, wikifactory, legacy, manager, goal, semantic, tool, cooperative, layers, healthier, defining, quantitative, knowledge, web, text, learning, techniques, computer, supporting, science, machine, documents, information, system |
| $c_{54}^{2006}$ | ontologies, language, query, specification, knowledge, manager, semantic, pure, capturing, data, search, keyword, layers, keyword-based, hybrid, architecture, spreadsheet, web, ie, application, information, modelling, approach, algorithm, using, methodic, retrieval, service, system, structures |

Table 4: Characterising keywords of Infomap community 54. The keywords are listed in their original, i.e., non-stemmed, form.

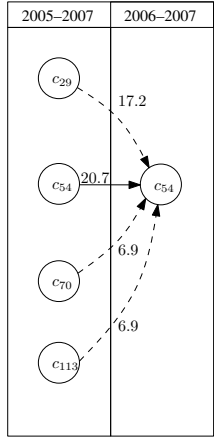

Figure 10: Main ancestors of community 54 in 2006

| time step | $\mathcal{S}$ | $\mathcal{T}$ | $\mathcal{A}$ |
|---|---|---|---|
| 2005 | 20 | 0.6107 | 0.81752 |
| 2006 | 29 | 0.42416 | 0 |

Table 5: Size $\mathcal{S}$, topic drift $\mathcal{T}$ and author entropy $\mathcal{A}$ of community 54 in 2005 and 2006

the SW and IR communities (see Figure 11(b)). Since 2004, there was not any IR-related keyword among the characterising keywords of community 5. Therefore, whereas community 5 kept its focus on the core SW-related topics, it largely participated in the formation of a new interdisciplinary community. This community, despite of still being focused mainly on SW-related themes, has functioned since then as a mutual intermediary between SW and IR communities. This hypothesis is supported by the high betweenness $\mathcal{B}$ value, especially in contrast with the betweenness in 2004 or with the average betweenness for the entire network (see Table 6). Note that even though the betweenness of community 0 in 2007 was even higher ($\mathcal{B} \doteq 2660$), this community was concentrated on core IR topics, and thus may not be perceived as an intermediary community between IR and SW camps. The analysis of content is thus a very valuable enhancement over the purely structural one in this case. Further analysis of the ancestors of community 15 led to the conclusion that, despite it emerged as an ancestor of IR-related community

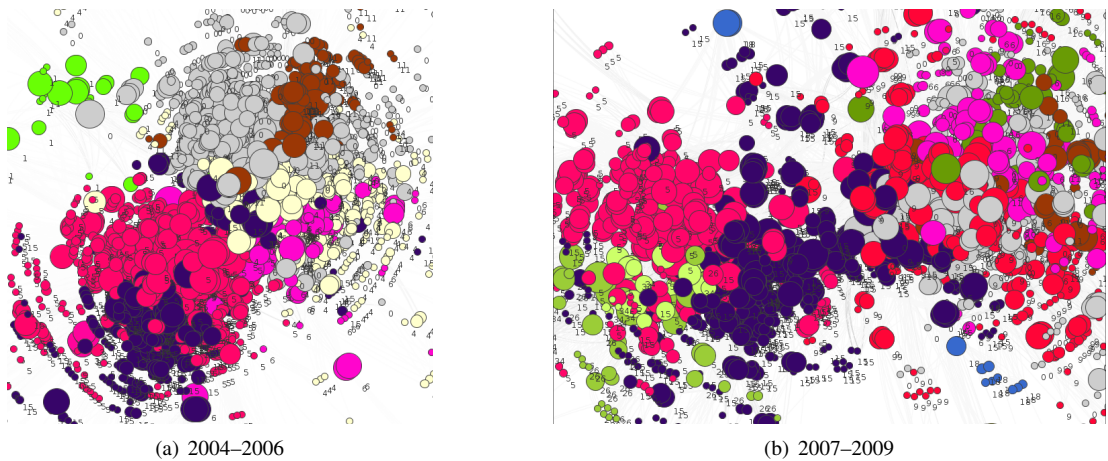

(a) 2004–2006  (b) 2007–2009

Figure 11: Community 5 "semantic web" (red—left side), "information retrieval" communities 0, 4, 6 and 9 (grey, beige, pink and red—right side, respectively) and their intermediary community 15 (violet)

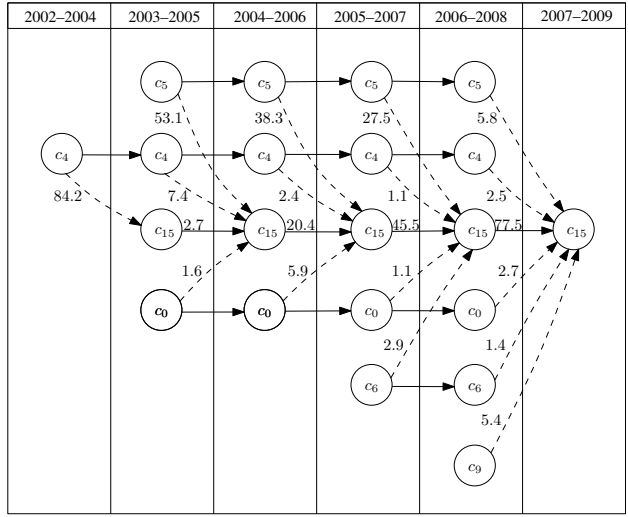

Figure 12: Main ancestors of intermediary community 15

4 and despite is was influenced by other IR-related communities 0, 6 and 9, it had been mainly formed by the semantic web community 5—especially in 2003 (by 53.1%), 2004 (by 38.3%) and 2005 (by 27.5%). Therefore, an effort to establish this interdisciplinary collaboration came mainly from the SW camp.

The community topic change measure $\mathcal{P}_\mathcal{C}$ helped to identify interesting and relevant events. However, these changes were rather only one aspect of the whole related cross-community dynamics and further interpretation using other measures was inevitable. This was particularly the case for Louvain community 15, which was detected by more than just one community topic evolution measure. But, the nature of this community, i.e., its intermediary role, was revealed by deeper analysis backed by visualisation and life-cycle measures. Average betweenness $\mathcal{B}$, author entropy $\mathcal{A}$ and *ancestor* measures were thus very helpful to gain a deeper understanding of changes of community topic. Other life-cycle measures like $\mathcal{T}$, $\rho$ or $\mathcal{H}$ were not very informative in this context.

|  | **2004–2006** | | | **2007–2009** | | |
|---|---|---|---|---|---|---|
|  | $\mathcal{S}$ | $\langle\langle\mathcal{B}\rangle\rangle$ | $\sigma^2_\mathcal{B}$ | $\mathcal{S}$ | $\langle\langle\mathcal{B}\rangle\rangle$ | $\sigma^2_\mathcal{B}$ |
| $c_{15}$ | 444 | 1591.01659 | $1.73509\times10^7$ | 445 | 2535.02 | $5.44681\times10^7$ |
| entire network | 2776 | 2066.70764 | $3.37936\times10^7$ | 2190 | 2192.85117 | $3.60915\times10^7$ |

Table 6: Size $\mathcal{S}$, average author betweenness $\mathcal{B}$ and its variance of community 15 and the entire network in 2004 and 2007

# 7 Discussion

In light of the results, this section further discusses the general features of community topic evolution measures, implications of the chosen matching method, some noticeable characteristics of the community-detection algorithms used and finally it wraps-up some of the community life-cycle events observed.

## 7.1 Community Topic Evolution Measures Revised

One might expect that the higher the values of the measures defined in Section 4.4.2 are, the stronger is the underlying community shift, merge or topic change. However, particularly very strong shifts $\mathcal{P}_\mathcal{S} \to 1$ identified in this way were associated with newly emerged communities, which disappeared in the next time step. Further analysis showed that many of them did not even dissolve to other clusters, but they disappeared completely from the network in following time-steps. These communities had usually very different yet coherent topics, and thus we assume that they might have been the initial sources of new topics or even research streams. Another possible explanation is that these communities ceased to exist entirely in the scientific world, which is hard to prove or reject, though. To explain this phenomenon, a larger data set that is not constrained to the a-priori chosen research fields is needed. Only then we can track the evolution of those communities after their emergence.

The community shift/merge measure $\mathcal{P}_{\mathcal{S}/\mathcal{M}}$ detected only one significant event, which may be caused by a relatively rare occurrence of this sort of events, or by the measure itself. Namely, if only a part of the community merges with a topically distinct community, this measure is not capable to detect it. We will have to revise measures for community shifts and merges and extend them by measures for actual *community split*.

Matching of communities by their Jaccard coefficient led in some cases to the situation that two communities were matched in spite of the relatively low Jaccard coefficient. One common approach is to use a matching threshold as in [14], but this may discard cases where the community does not dissolve, but is rather influenced by new members joining it. On the other hand, matching of communities just by means of an overlap measure like the Jaccard coefficient sometimes led to high values of $\mathcal{P}_\mathcal{C}$. This may be perceived as a false-positive result depending on how the community is defined: whether the community keeps its identity even though it has been significantly influenced by other communities or not. In case of the latter interpretation, the matching process may be improved by employing the topical similarity in a way that only similar communities are matched. This would, however, cause the current measure for community topic change to stop working, because changes of community topic would be discarded early in the process of community matching.

## 7.2 Clustering Algorithms

Using the proposed methodology, we were able to identify desired cross-community phenomena among clusters obtained by both Louvain and Infomap algorithms. Some basic comparison was provided in Section 4.5. The obvious difference between these two algorithms is that Louvain produced clusters of bigger size. On the one hand, this led to smaller amount of result sets of community topic evolution measures, which could be processed and interpreted in less time. On the other hand, this was probably rather a coarse-grained picture of the community structure, and thus some of the interesting phenomena might have been missed. Infomap sometimes produced very small clusters with $\mathcal{S} < 10$, which very often dissolved immediately. As this method does not suffer a resolution limit as modularity-optimisation methods do [9], this algorithm was able to identify even small communities featuring rare phenomena like community shift/merge. Thus, depending on the aims of the researcher and the analysis, either algorithm will provide clustering suitable for the analysis of the cross-community phenomena: Louvain being particularly suitable for coarse-grained quick insights, whereas Infomap suitable for fine-grained deeper analysis. This also shows that it is worthwhile to investigate more community-detection algorithms, particularly ones producing overlapping communities and evolutionary clustering methods.

### 7.3 Community Life-Cycles

The most frequent phenomenon described was community shift. Especially the pattern where one big community produced a couple of new communities with more-or-less different topic was quite common. These shifts thus were part of the community specialization process of their common ancestor. This was the case of mutual relationship of Infomap community 9 and 99 described in Section 6.1. This pattern may be characterised by a decreasing size of the main and general community (e.g., community 9), while its cluster content ratio increases in parallel.

The process of stabilisation of a newly emerged community was common as well. These communities had often a weak structure (low relative density, high entropy) and/or topic (low topic drift) at the beginning, while characteristics like size, topic drift or relative density were increasing in subsequent time-steps. These measures thus seem to be good candidates for further automated analysis of important events in the life-cycle of a community.

In detection of intermediary communities like Louvain community 15 (see Section 6.3), the average vertex betweenness, author entropy and *ancestor* relation have shown to be informative. Using solely average author betweenness, we identified another intermediary community in Infomap clustering, whose betweenness peaked suddenly at one point. The position right between IR and SW camps, identified by topical analysis, supported the hypothesis of an intermediary. Therefore, this measure may be a valuable complement to community topic evolution measures in that regard.

In general, the many cases we found for some of the investigated phenomena suggest that they are part of a usual community life-cycle. Moreover, there seem to be general mechanisms that shape the evolution of scientific communities, which can be identified by (a combination) of some of the proposed life-cycle measurements. Further investigation, particularly on a larger data set, will have to show if this is really the case. The appropriate combination of topological and topical analysis seems to be inevitably for the accurate detection, understanding and prediction of a community's actual life-cycle.

## 8 Conclusion and Future Work

We presented a general methodology for analysing community dynamics, uniquely combining *topological* and *topical* analysis and supported by special visualisation techniques. In this light, we focused on cross-community effects and tried to explore typical life-cycles of scientific communities. Exemplary, the methodology was applied to the co-citation network of scientists from two related research fields, IR and SW. Three *community topic evolution measures* tailored for identifying phenomena like community shift, merge/shift and change of topic were proposed and successfully assessed. Community shift and community topic change were detected quite commonly, which suggests that they are part of many community life-cycles. Community shift/merge was detected rarely, which either means we have to improve the used measure or that this is simply a rare phenomenon.

To improve the topic evolution measures and to overcome some of their limitations, we proposed *life-cycle measures* characterising the states and evolution of communities. An assessment showed that average vertex betweenness, cluster quality, relative density and topic drift offered very valuable insights into the phenomena revealed by community topic evolution measures. We intend to cluster the community life-cycles by the characteristic events expressed by all the aforementioned measures, which we expect to provide an automated way of extracting life-cycle taxonomies.

In addition to the raw analyses, visualisation proved to be a valuable way for formulating first hypotheses. We used a new tool implemented specifically for the tracking of dynamic cross-community phenomena. Particularly the identification of *intermediary communities* was simplified by this. However, the automated construction of evolution diagrams proved to be a challenging task. In this light, the joint visualisation of these diagrams and tag clouds is one of the promising tasks for future work. This will further increase the benefits we gain from the extracted meta-data.

The combination of content and structural analysis of communities enabled a high-quality analysis of community dynamics and further allowed us to assess the quality of clustering methods. However, the focus of this work was on cross-community phenomena. Nevertheless we believe that this original approach to cluster quality assessment is a fertile ground for future research and worth further investigation.

As the pluggable design of the framework brings the flexibility of choice regarding the community-detection method, we plan to use other algorithms—especially those specifically tailored for co-clustering of both content and objects [13]. Further, we will extend the whole work to a larger data set. Long-term plans also involve the investigation of different author relationships and a closer look at the micro level of individual community members.

## Acknowledgement

We thank to our colleagues Georgeta Bordea and Dr. Paul Buitelaar from the Unit for Natural Language Processing for their help in extracting and processing the meta-data. The material presented in this work is based upon works jointly supported by the Science Foundation Ireland under Grant No. SFI/08/CE/I1380 (Lion-2) and under Grant No. 08/SRC/I1407 (Clique: Graph & Network Analysis Cluster).